\documentclass[twocolumn,preprintnumbers,amsmath,amssymb,aps,prb]{revtex4}
\usepackage{graphicx}
\begin{document}

\title{Peak Effect, Melting, and Transport in Skyrmion Crystals}
 
\author{C. Reichhardt$^{1,2}$ and C.~J.~O. Reichhardt$^{1,2}$}
\affiliation{$^{1}$Theoretical Division and Center for Nonlinear Studies,
Los Alamos National Laboratory, Los Alamos, New Mexico 87545, USA}
\affiliation{$^{2}$Center for Nonlinear Dynamics, Los Alamos National Laboratory, Los Alamos, New Mexico 87545}

\date{\today}

\begin{abstract}
We numerically examine the transport of skyrmions driven over
weak random quenched disorder using a modified Thiele approach that
includes
the thermal softening of skyrmion pairwise interactions
introduced by Wang {\it et al.}, Phys.~Rev.~Appl.~{\bf 18}, 044024 (2022).
The depinning transition is elastic at low temperatures
but becomes plastic
with a reduced threshold at higher temperatures
due to the competition between thermal creep and thermal softening,
indicating a temperature-induced order to disorder transition into
a glass state.
The resulting non-monotonic critical depinning forces and
crossing of the velocity-force curves
are similar to what is observed
in the peak effect for type-II superconducting 
vortex lattices, where the softening of vortex-vortex interactions with
temperature leads to an order-disorder transition.
For low skyrmion densities the peak effect is absent
since the system is always in 
a disordered state. 
We map the dynamical phase transition
as a function of temperature, density,
drive, and materials parameters, and show that the signatures are similar to
those of the superconducting peak effect.
Our results are consistent with
recent experiments.

\end{abstract}

\maketitle

\vskip 2pc

\section{Introduction}
Skyrmions are particle-like magnetic textures
that can form in chiral magnets
\cite{Muhlbauer09,Yu10,Nagaosa13,EverschorSitte18}. They
have been attracting growing 
attention as an increasing number of  materials are
being found that support skyrmions,
and a variety of possible applications have been proposed for using
skyrmions for  memory and novel computing \cite{Fert17,Wang22}.  
Skyrmions have many similarities to vortices in
type-II superconductors \cite{Blatter94},
including the fact that
they have repulsive pairwise interactions \cite{Lin13,Brearton20},
form triangular lattices \cite{Muhlbauer09,Yu10,Nagaosa13},
can be set into motion with 
an applied current \cite{Nagaosa13,EverschorSitte18,Schulz12,Yu12,Iwasaki13},
and
can interact with and be pinned by quenched disorder
so that there is a critical
driving force required to set them in motion
\cite{Schulz12,Iwasaki13,Reichhardt15,Woo16,Legrand17,Reichhardt22a,Gruber22}.
In many 
skyrmion systems, thermal fluctuations are also relevant,
and skyrmions have been
shown to exhibit
diffusion \cite{Zazvorka19,Zhao20}, creep behavior
\cite{Lin13,Reichhardt22a,Reichhardt18a,Luo20,Litzius20},
skyrmion lattice melting \cite{Huang20,Balaz21}, 
and skyrmion glass phases \cite{Hoshino18}.
An important difference between skyrmions and superconducting
vortices is that
the skyrmions have
a strong gyrotropic component to their dynamics due to the Magnus
force, which causes the skyrmions to move at an angle 
with respect to an applied driving force
\cite{Nagaosa13,EverschorSitte18,Lin13,Iwasaki13,Reichhardt15,Legrand17,Reichhardt22a,Litzius20,Jiang17,Litzius17,Juge19}.
This gyrotropic 
term also strongly affects how skyrmions interact with pinning
\cite{Reichhardt15,Legrand17,Reichhardt22a}. 

One of the most interesting effects observed for superconducting
vortex systems is the peak effect, in which the critical depinning
force is nonmonotonic and passes through
a peak with increasing magnetic field
\cite{DeSorbo64,Pippard69,Wordenweber86a,Bhattacharya93,Kwok94,Banerjee00}.
Here, the depinning threshold initially
decreases with increasing field as the vortex density increases
since the
vortex lattice is becoming stiffer;
however, just below the upper magnetic field $H_{c2}$ above which
superconductivity is destroyed,
the depinning threshold shows a rapid increase followed by
a downward plunge,
leading to a peak
in the depinning force
\cite{DeSorbo64,Pippard69,Wordenweber86a,Bhattacharya93}. 
It was argued that the pairwise vortex-vortex interactions
are modified and reduced
at high fields,
leading to a softening of the vortex lattice and a reduction
of its shear lattice
\cite{Pippard69}.
A softer lattice can
more readily adapt to a random pinning landscape and is
better pinned than a stiff lattice.
The velocity-force curves also
change shape across the peak effect \cite{Bhattacharya93},
producing a crossing of the curves in which
the depinning threshold is lowest
above the peak effect but the flow velocity is lowest
below the peak effect
\cite{Higgins96}. The vortex peak effect also occurs
for fixed vortex density as function of increasing 
temperature
\cite{Kwok94,Paltiel00a,Ling01,Troyanovski02,Mohan07,ToftPetersen18}.
Again this is counterintuitive
since in general for a particle-based
system coupled to quenched disorder, an increase in thermal
fluctuations should cause
the particles to jump more readily out of the wells,
resulting in a monotonic decrease of the critical depinning
force $F_{c}$
with increasing temperature $T$ \cite{Fisher98,Reichhardt17}.
In the vortex case, the  vortex-vortex
interactions are also modified by the temperature,
leading to a softening of the vortex lattice 
with increasing temperature.

It has been argued that the superconducting
peak effect is
produced by a transition from an
ordered lattice to a disordered phase that occurs when 
the vortex lattice becomes soft enough for topological defects to
proliferate due to the pinning,
strongly reducing or breaking down the elasticity of the
lattice \cite{Bhattacharya93}.
In this sense, the peak effect can also be viewed
as a melting transition in the presence of quenched disorder
\cite{Kwok94,Banerjee00}. 
Neutron scattering \cite{Ling01,Gammel98}
and imaging experiments \cite{Troyanovski02,Ganguli15} have shown
transitions from an
ordered vortex state below the peak effect to disordered structures 
at and above the peak effect. In the
thermal peak effect, as $T$ is further increased above the peak,
$F_{c}$ decreases again due to increasing thermal hopping in the disordered 
state. 
The thermal peak effect can be regarded as resulting from
a competition between
the reduction of the depinning threshold by increasing temperature
and the reduction of the pairwise vortex-vortex interactions that
tends to increase the depinning force.
The sudden increase in the
critical force then corresponds to a transition from
ordered elastic depinning to disordered plastic depinning.
Such a change from elastic to plastic depinning
will also alter
the shape of the velocity-force curves, which have the form 
$v = (F_{D} -F_{C})^{\beta}$ with $\beta < 1.0$ in the elastic
depinning regime and
$\beta > 1.0$ in the plastic depinning regime \cite{Reichhardt17}.

Noise fluctuations can also be used
to probe the peak effect.
An ordered lattice moves with low noise,
but near the peak effect 
the vortex lattice breaks apart and depins plastically,
producing large
large noise,
while above the peak the flow becomes
more fluid-like and the noise is low again 
\cite{Marley95}.
Since the peak effect separates two different phases, 
various memory and metastable effects occur
when there is a coexistence of the 
ordered and disordered phases \cite{Paltiel00a,Henderson96,Banerjee99}. 
Another observation is that the peak effect 
does not occur in strongly pinned samples
where the vortices are always in a disordered state \cite{Banerjee00}.
In general, for systems with weaker pinning
the peak effect occurs at higher temperatures, indicating that
as the pinning strength decreases, the
lattice must be softer in order to permit
the order to disorder transition to occur
\cite{Banerjee00}.
There can also be a jump up in the critical depinning force at
low fields
as the vortex density is decreased, which
occurs when the vortices become so far apart that the vortex lattice
shear modulus drops to a low value
and the
system becomes disordered \cite{Ghosh96}.
Simulations with modifications
to the vortex-vortex interactions
have produced a peak in the depinning force \cite{Tang96}. 

An open question is whether a skyrmion lattice
can also show a peak effect
phenomenon similar to that found in the superconducting system,
and whether there would be similar changes in
the transport properties across the order to disorder transition. 
Recently Wang {\it et al.} \cite{Wang22a}
numerically studied the interactions $U(T)$
between 
two skyrmions as a function of temperature
and found that the interactions decrease
approximately exponentially with increasing temperature.
They proposed that the functional form
$U(T) \propto \exp(-T/\lambda^*$),   
where $\lambda^*$ is a material parameter,
could
be used to introduce temperature dependence of the interaction
into a particle-based
model of skyrmions.
An interaction of this type
suggests that if a skyrmion lattice is in the
presence of pinning,
under increasing temperature
the interactions
sharply decrease and
the lattice softens, making it possible for a transition to
occur from an ordered to a disordered lattice.

Interestingly, in the literature \cite{Schulz12}
it was found that the depinning threshold for skyrmions in MnSi increases
with temperature,
consistent with the idea that
the skyrmion lattice has softened and become better pinned.
Luo {\it et al.} \cite{Luo20} used ultrasound spectroscopy to
examine a skyrmion lattice in MnSi and
measure the depinning force.
They found a regime where the critical depinning current
decreases with increasing $T$ followed by an upturn
for further increases in $T$  that is correlated with a softening
of the skyrmion lattice.
This was interpreted as
evidence for a skyrmion peak effect similar to that
found in the superconducting vortex systems.

In this work we consider a particle based model for
skyrmions interacting with pinning where we
add thermal fluctuations and modify the skyrmion pairwise
interactions according to the exponential decrease with $T$ proposed
by 
Wang {\it et al.} \cite{Wang22a}. 
We focus on parameters where the
skyrmions form a lattice that depins elastically at $T = 0$.
As the temperature increases,
we find that 
the depinning threshold initially decreases due to
increased creep;
however, when the pairwise interactions become
weak enough,
the skyrmion lattice disorders and
there is a pronounced 
increase in the depinning threshold.
As the temperature is further increased, the depinning 
threshold decreases again due to
increased thermal hopping and eventually goes to zero when the  
skyrmions form a liquid.
The result is the appearance of a
peak effect in the depinning force
similar to that found for superconducting vortices. 
At low skyrmion densities or in regimes where the 
skyrmions are always disordered even for $T=0$,
there is no peak effect.
We also find that the
velocity-force curves change concavity across the peak effect,
resulting in a crossing of the curves.
Above the peak effect temperature $T_p$
where the depinning threshold is low, the 
velocity in the flow regime is lower than that found for 
the same drive 
in the lower temperature elastic system. 
We also show that the skyrmion Hall angle is nonmonotonic and
changes across the peak effect regime.
Under an increasing external drive at temperatures
above $T_p$ but below the melting temperature $T_m$,
the system can undergo dynamical reordering similar
to that found for superconducting
vortices
\cite{Bhattacharya93}. The drive at which reordering occurs
diverges as
$T_{m}$ is approached, 
in agreement with the predictions of Koshelev and Vinokur for driven 
superconducting vortices \cite{Koshelev94}.  
We construct a dynamic phase diagram and show that it is similar
to the diagram observed for superconducting vortices
\cite{Bhattacharya93,Hellerqvist96}, with
pinned lattice, 
pinned glass, plastic flow, and moving crystal
regimes.
For constant drive well above the depinning threshold,
we find 
that as the temperature increases, 
the velocity and skyrmion Hall angle both drop
across the elastic to plastic phase transition,
but increase at higher temperature and saturate above $T_m$ to the values
expected for a pin-free system.
Our results could be tested with transport, imaging, neutron scattering and
noise measurements in skyrmion
samples that show a lattice regime. 

\section{Simulation}
We work with a particle based model for skyrmions
employing a modified Thiele equation 
approach that has been used in previous studies of skyrmion ordering,
depinning, transport,
and thermal effects
\cite{Lin13,Reichhardt15,Reichhardt22a,Reichhardt18a,Brown19,Stidham20,Zhou21}. 
We consider a two-dimensional (2D)
system of size $L \times L$
with periodic boundary conditions in the $x$ and $y$ directions
containing $N_{s}$ skyrmions and $N_{p}$ pinning sites.  
The  equation of motion for a single skyrmion $i$ is
\begin{equation} 
\alpha_d {\bf v}_{i} + \alpha_m {\hat z} \times {\bf v}_{i} =
	{\bf F}^{ss}_{i} + {\bf F}^{sp}_i +   {\bf F}^{D}_{i} + {\bf F}^{T} \ .
\end{equation}
The skyrmion velocity is 
${\bf v}_{i} = {d {\bf r}_{i}}/{dt}$
and $\alpha_d$ is the damping term
that aligns the skyrmion velocity in the direction of drive.
There is also a gyrotropic or Magnus force
component, with coefficient
$\alpha_{m}$,
that generates a velocity perpendicular to the net forces acting
on the skyrmion.
The repulsive skyrmion-skyrmion interaction force is
${\bf F}_{i}^{ss} = \sum^{N_s}_{j=1}A_{s}K_{1}(r_{ij}){\hat {\bf r}_{ij}}$,
where $r_{ij} = |{\bf r}_{i} - {\bf r}_{j}|$ is
the distance between skyrmions $i$ and $j$
and $\hat {\bf r}_{ij}=({\bf r}_i-{\bf r}_j)/r_{ij}$.
The coefficient $A_s$ is material dependent.
The first order Bessel function form $K_{1}(r)$
of the interaction was obtained from a fit to a continuum model \cite{Lin13},
and it decreases exponentially at longer range.
Imaging  
studies have found similar skyrmion interactions
that decay exponentially at longer range \cite{Ge23}.
The pinning force
${\bf F}_{i}^{sp}=\sum_{k=1}^{N_p}(F_p/r_p)\Theta(r_{ik}^{(p)}-r_p){\bf \hat{r}}_{ik}$
arises from non-overlapping attractive parabolic wells of maximum
range $r_{p}$ and maximum strength $F_{p}$,
where $r_{ik}^{(p)}=|{\bf r}_i-{\bf r}_k^{(p)}|$ is the distance
between skyrmion $i$ and pin $k$ and
${\bf \hat{r}}_{ik}^{(p)}=({\bf r}_i-{\bf r}_k^{(p)})/r_{ik}^{(p)}$.
This is the same form of substrate
used in previous particle-based models to study
skyrmion depinning \cite{Reichhardt15,Reichhardt22a}. 
The driving force ${\bf F}^{D}=F_D{\bf \hat{x}}$
is produced by an applied spin-polarized
current. 
The thermal term ${\bf F}^{T}$ is modeled by Langevin kicks with
the properties
$\langle F^T\rangle=0$ and
$\langle F_i^T(t)F_j^T(t^\prime)\rangle=2\eta k_BT\delta_{ij}\delta(t-t^\prime)$.

In the absence of interactions with other skyrmions or pinning sites,
a skyrmion  
moves at an
intrinsic Hall angle
of $\theta^{\rm int}_{sk} = \arctan(\alpha_{m}/\alpha_{d})$ with respect to the
applied driving force.
In the overdamped case, $\theta^{\rm int}_{sk} = 0^\circ$.
We take $\alpha^2_{d} + \alpha^2_{m} = 1.0$, and
throughout this work we fix
$\alpha_{d} = 0.866$ and $\alpha_{m} = 0.5$,
giving an intrinsic skyrmion Hall angle of
$\theta^{\rm int}_{sk}=-30^\circ$
that is well within the range of experimental values.
We measure the average skyrmion velocity both parallel,
$\langle V_{||}\rangle=\langle N_S^{-1}\sum_i^{N_s}{\bf v}_i \cdot {\bf \hat{x}}\rangle$, and
perpendicular,
$\langle V_{\perp}\rangle=\langle N_s^{-1}\sum_i^{N_s}{\bf v}_i \cdot {\bf \hat{y}}\rangle$, to the applied drive, where each quantity is averaged over
$1 \times 10^{5}$ simulation time steps.
From these measurements we obtain
the observed
skyrmion Hall angle $\theta_{sk} = \arctan(\langle V_{\perp}\rangle/\langle V_{||}\rangle)$,  
which is affected by
interactions with the pinning landscape, as
shown in both simulation and experiment,
and generally increases with increasing drive before
saturating to the clean value at
high drives well above depinning
\cite{Reichhardt15,Reichhardt22a,Jiang17,Litzius17,Juge19}. 

Wang {\it et al.} \cite{Wang22a}
proposed that the
strength $U(T)$ of the pairwise skyrmion-skyrmion interactions
decreases exponentially with temperature according to the
approximate form
$U(T) \propto \exp(-T/\lambda^*)$, where $\lambda^*$ is  a material
parameter. 
Here we consider the interaction of two thermal effects.
As the temperature $T$ increases, the magnitude $F^T$ of the thermal
fluctuations also increases, which increases the diffusion of the
skyrmions.
At the same time,
the skyrmion-skyrmion interaction coefficient varies as
$A_s=A_s^0\exp(-\kappa T)$, causing the skyrmion lattice to soften with
increasing temperature.
We focus on a system of size $L = 36$
with $r_{p} = 0.25$, $N_{s} = 450$,
a skyrmion density of $n_s=0.4$,
$N_p=672$, a pinning density of $n_p=0.6$,
$F_p=0.1$, $A_s^0=3.0$, and $\kappa = 2.0$, but 
in later sections of the paper
we also consider different
skyrmion densities and vary $\kappa$ over the range $\kappa =0$ to $3.0$.   
 
\section{Results}
\begin{figure}
\includegraphics[width=3.5in]{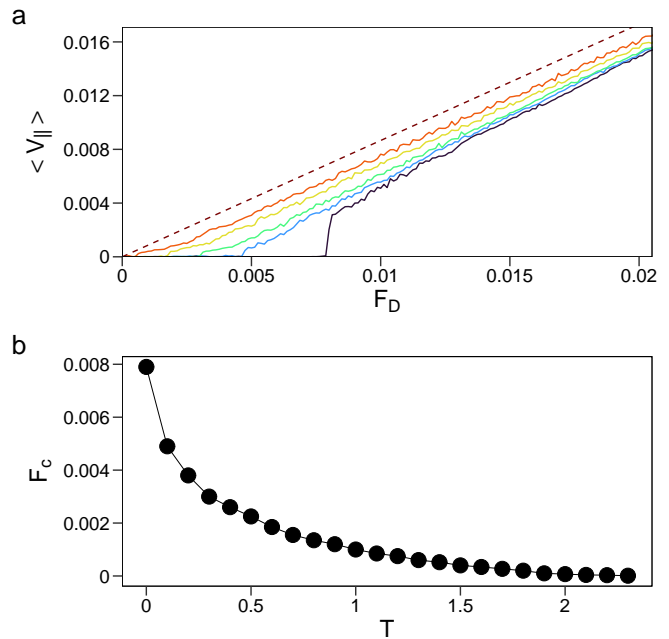}
\caption{ 
(a) The skyrmion velocity $\langle V_{||}\rangle$ in the direction of
the applied drive
vs $F_{D}$
in a system with
fixed skyrmion-skyrmion interaction strength
$A_{s} = A_s^0$
at
temperatures $T = 0$, 0.1, 0.2, 0.6, and
$1.4$, from bottom to top.
The dashed line is the expected
response in the absence of pinning.
(b) The
critical depinning force 
$F_{c}$ vs $T$ for the system in 
(a). The depinning threshold
decreases monotonically with increasing $T$.
}
\label{fig:1}
\end{figure}

We first consider the effects
on the depinning behavior of independently varying the
thermal fluctuations and the skyrmion-skyrmion interactions.
We focus on a system with skyrmion density $n_{s} = 0.4$,
$F_{p}= 0.1$, and $A_s=A_{s}^0$.
For these parameters,
at zero temperature the skyrmions form
a triangular lattice that depins elastically without
the generation of topological defects.

In Fig.~\ref{fig:1}(a) we plot the velocity-force curves
$\langle V_{||}\rangle$ versus $F_{D}$ for
temperatures of $T = 0$, 0.1, 0.2, 0.6, and  $1.4$. 
Here the skyrmion-skyrmion interaction is not modified as $T$
increases; only the magnitude of the thermal fluctuations increases.
The dashed line shows the expected response in a pin-free
sample with $F_p=0$.
We find that the
depinning threshold decreases monotonically with increasing $T$, and 
above depinning,
$\langle V_{||}\rangle$ for a fixed value of $F_D$ also decreases
monotonically.
Figure~\ref{fig:1}(b) shows
that the critical depinning force $F_{c}$
decreases 
with increasing $T$ and drops to zero
for $T > 2.0$,
well below
the thermally induced melting temperature
of $T_m = 3.0$.  
The critical force obtained from the perpendicular velocity
$\langle V_{\perp}\rangle$ (not shown) exhibits
similar behavior, and the
drive dependent skyrmion Hall angle
saturates at higher drives
as discussed
in previous work \cite{Reichhardt18a}. 

\begin{figure}
\includegraphics[width=3.5in]{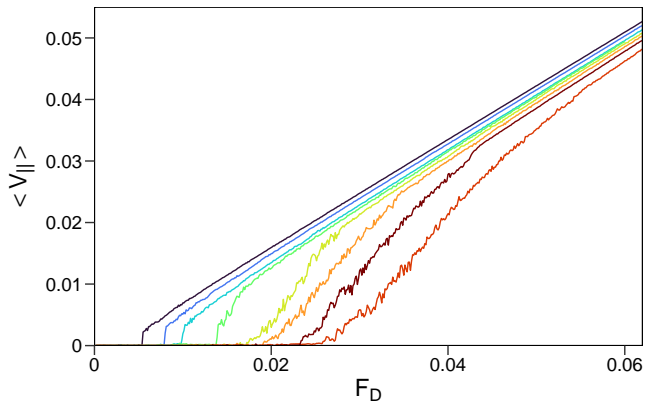}
\caption{ $\langle V_{||}\rangle$
vs $F_{D}$ for the 
system in Fig.~\ref{fig:1} at $T = 0$
for modified skyrmion-skyrmion interaction strengths of
$A_s  = 6.0$, 3.0, 1.75, 1.5, 1.25, 1.0, 0.75
and $0.5$, from top to bottom.
The depinning threshold monotonically increases with decreasing $A_s$. 
The depinning
transition is elastic for $A_s> 1.25$ and plastic for
$A_s \leq 1.25$.  
}
\label{fig:2}
\end{figure}

Next we introduce thermal modification of the skyrmion-skyrmion
interactions but eliminate the thermal fluctuations so that the
Langevin kicks have a temperature of $T=0$.
In Fig.~\ref{fig:2} we show some representative
velocity-force curves for
decreasing values of $A_s  = 6.0$, 3.0, 1.75, 1.5, 1.25, 1.0, 0.75,
and $0.5$. The depinning force $F_c$
increases monotonically 
with decreasing $A_s$ due to the softening of the skyrmion-skyrmion
interactions.
For $A_s = 6.0$, 3.0, 1.75, and 1.5,
the depinning transition is elastic and the skyrmions maintain
the same neighbors as they depin,
while for $A_s = 1.25$, 1.0, 0.75, and 0.5,
the depinning is plastic and the skyrmion lattice breaks apart
into river-like regions of flow.
We also find that
the shape of 
the velocity-force curves changes across the
transition from elastic to plastic depinning.

\begin{figure}
\includegraphics[width=3.5in]{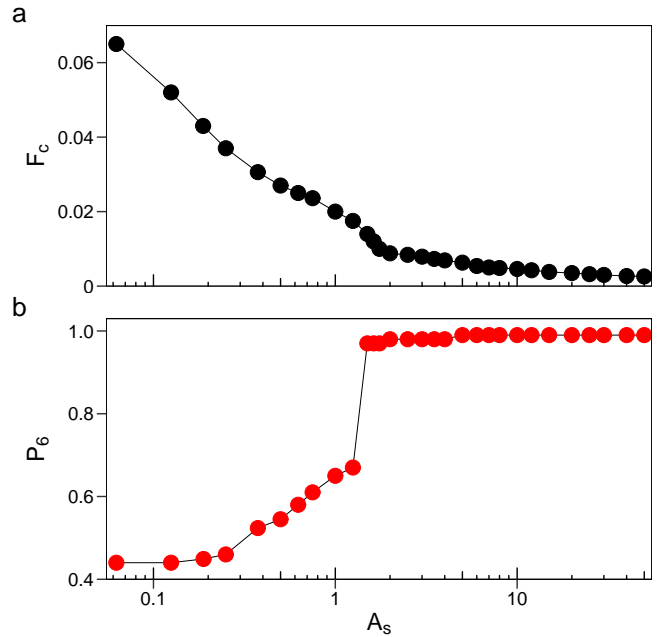}
\caption{ 
(a) $F_{c}$ vs $A_s$ for the system in Fig.~\ref{fig:2} with $T=0$.
(b) The corresponding fraction $P_6$ of sixfold-coordinated
skyrmions vs $A_s$.
As $A_s$ decreases, $F_c$ increases, with a more rapid increase appearing
at the order to disorder transition near $A_s=1.25$.  
}
\label{fig:3}
\end{figure}

In Fig.~\ref{fig:3}(a) we plot $F_{c}$ versus $A_s$
for the $T=0$ samples from Fig.~\ref{fig:2}, while
Fig.~\ref{fig:3}(b) shows the
corresponding fraction $P_6$ of 
sixfold-coordinated
skyrmions versus $A_s$.
Here $P_{6}=N_s^{-1}\sum_i^{N_s}\delta(z_i-6)$, where $z_i$ is the
coordination number of skyrmion $i$ obtained from
a Voronoi construction.
For $A_s > 1.25$, $P_6 \approx 1.0$,
the depinning is elastic, and $F_{c}$ decreases
slowly with increasing $A_s$.
For $A_s \leq 1.25$,
$P_{6}$ drops with decreasing $A_s$ as the system becomes disordered, and
there is a relatively rapid increase in $F_{c}$
since the disordered skyrmion lattice
is better pinned.
As $A_s$ decreases further,
$F_{c}$ continues to increase, and the maximum
critical force of $F_c/F_p=1.0$ occurs when $A_s=0$ (not shown).
The onset of plastic depinning with decreasing $A_s$ can be
viewed as representing
the first part of the peak effect
in which 
the topologically disordered lattice
softens and can
be better pinned than a clean elastic lattice.
As $A_s$ decreases further, however,
$F_{c}$ continues to grow,
and at $A_{s}=0$, $F_{c}=F_p$,
so there is no peak in $F_c$ at nonzero values of $A_s$.

\begin{figure}
\includegraphics[width=3.5in]{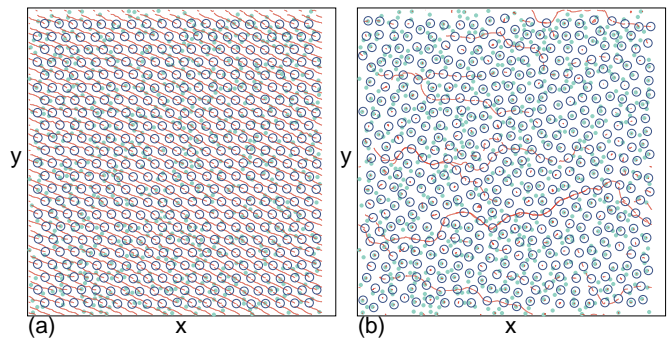}
\caption{
The pinning site locations (filled circles),
skyrmion locations (open circles), and trajectories (lines)
for the system in Fig.~\ref{fig:3} with $T=0$.
(a) At $A_s = 3.0$ just above 
depinning, the flow is elastic.
(b) At $A_s = 0.5$, the skyrmions are disordered
and the depinning is plastic. 
}
\label{fig:4}
\end{figure}

In Fig.~\ref{fig:4}(a) we illustrate the pinning site locations and skyrmion
positions and trajectories
for the system in Fig.~\ref{fig:3} at $A_s = 3.0$ just above depinning,
where the skyrmions form a lattice that moves as a rigid object.
It is also clear that the motion occurs at an angle
to the applied drive that is
associated with the skyrmion
Hall angle.
Figure~\ref{fig:4}(b) shows the same
system at $A_s = 0.5$ where the skyrmions 
are disordered and only some of the skyrmions are moving while the rest
of the skyrmions remain pinned.
These results show that
in the absence of thermal fluctuations,
softening the skyrmion lattice increases the depinning threshold,
with a rapid increase in the critical force $F_c$
at the order to disorder transition. 
This suggests that if thermal fluctuations are combined with
a weakening of the skyrmion-skyrmion interaction as temperature
increases,
the two effects will compete, resulting in the emergence of
a peak effect in the depinning threshold.

\section{Thermal modification of the Skyrmion-Skyrmion Interaction}

We next consider the case where thermal fluctuations are increased
at the same time that
the skyrmion-skyrmion interaction is adjusted for temperature
using the exponential
dependence proposed by Wang {\it et al.} \cite{Wang22a},
$A_s = A_s^0 \exp(-\kappa T)$.
We fix $\kappa = 2.0$ and 
set $A_s^0 = 3.0$, so that
at $T = 0$ the system forms a triangular solid that depins elastically,
as shown in Fig.~\ref{fig:3}.
For these parameters, in a pin-free system with
$F_{p} = 0$, 
a melting transition
occurs at $T_m = 1.125$ via the proliferation
of topological defects. 

\begin{figure}
\includegraphics[width=3.5in]{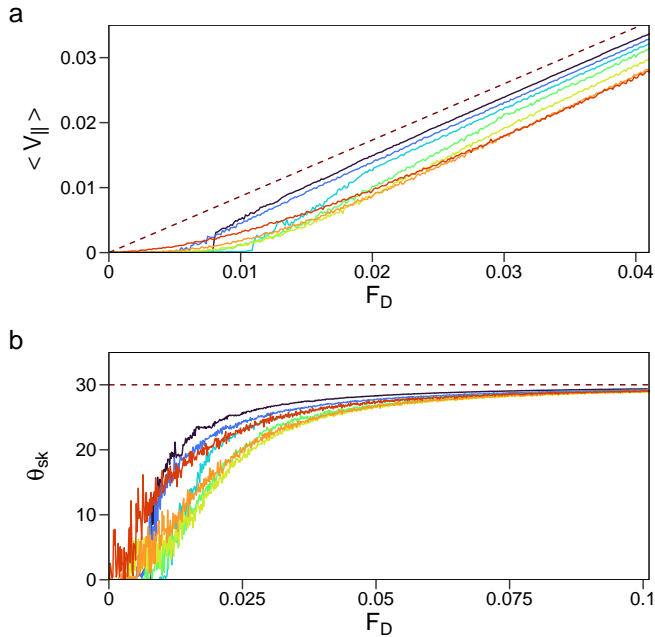}
\caption{
$\langle V_{||}\rangle$ vs $F_D$  
for a system in which we simultaneously increase $T$ while
modifying the skyrmion-skyrmion interaction $A_s$
with temperature according to the
relation proposed by
Wang {\it et al.} \cite{Wang22a}.
From top right to bottom right,
$T/T_m=0$, 0.26, 0.4346, 0.6086, 0.869, 1.3043, and 1.739,
where $T_m$
is the temperature at which a pin-free system with $F_{p} = 0$
melts.
The depinning threshold
initially decreases with increasing temperature, increases again
to reach its maximum value
near $T/T_m = 0.5$, and then deceases again for high $T/T_m$.
We also find a
crossing of the velocity-force curves.
(b) The corresponding skyrmion Hall angle $\theta_{sk}$ vs $F_{D}$.
Dashed lines indicate the expected response in the absence of pinning.
}
\label{fig:5}
\end{figure}

In Fig.~\ref{fig:5}(a) we plot
$\langle V_{||}\rangle$ versus
$F_{D}$ for
$T/T_m=0$, 0.26, 0.4346, 0.6086, 0.869, 1.3043, and 1.739,
while Fig.~\ref{fig:5}(b) shows the corresponding
skyrmion Hall angle $\theta_{sk}$ versus $F_D$. 
The system depins elastically for $T/T_m = 0$
and $T/T_m = 0.2$, and in this regime the depinning force
decreases with increasing $T/T_m$. 
The shape of the velocity-force curve changes
for $T/T_m = 0.4346$
and the depinning threshold increases since the depinning
transition becomes plastic.
For even higher $T/T_m$,
there is an increasing amount of creep in which
the velocity becomes finite at lower $F_{D}$. 
A crossing of the velocity-force curves occurs in Fig.~\ref{fig:5} since
the velocities at higher $F_D$ are lower for higher values of $T/T_m>0.4346$
compared to $T/T_m<0.4346$, even though the depinning threshold for
$T/T_m>0.4346$ is lower than that found for $T/T_m<0.4346$.
In the elastic depinning regime for $T/T_m < 0.4346$, 
$\theta_{sk}$ becomes finite just above depinning,
and it initially has a low value but increases
rapidly with increasing $F_D$.
The steps
that appear in the velocity
at depinning when $T/T_m = 0$ are caused by an alignment of the
angle of skyrmion motion
with the symmetry direction of the triangular
pinning lattice,
an effect that was predicted to lead to steps
in the velocity-force curves \cite{LeDoussal98}.
This directional locking was studied previously for skyrmions
\cite{Reichhardt19b}.
In the plastic depinning regime, the skyrmion Hall angle is generally lower.
The fluctuations of $\theta_{sk}$ are largest just above depinning since
the angle is calculated from a ratio of two velocities.
Figure~\ref{fig:5}(b) indicates that the skyrmion Hall angle is 
also non-monotonic as a function of temperature.
The lowest value of $\theta_{sk}$ appears for
$T/T_m = 1.0$,
and there is a crossing of the skyrmion Hall angle
curves that is similar to the crossing of the velocity-force
curves. The behavior of the skyrmion Hall angle
through the peak effect region
is discussed in greater detail in Section~V.

\begin{figure}
\includegraphics[width=3.5in]{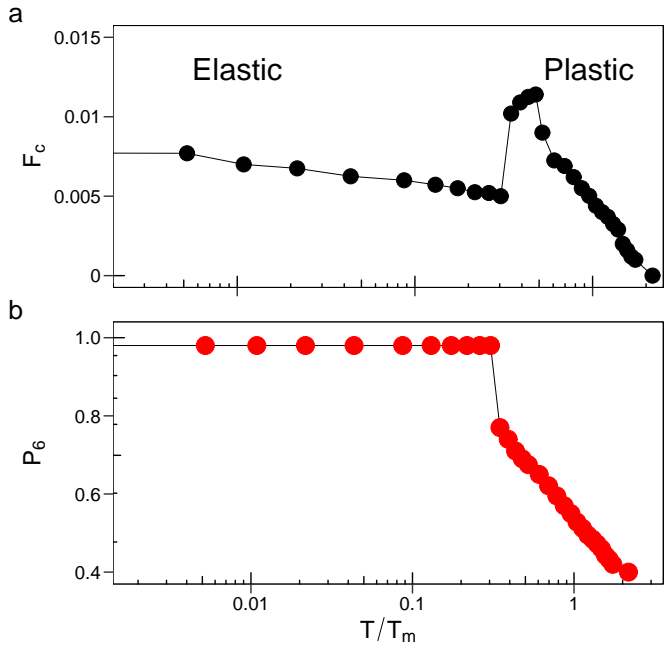}
\caption{(a) $F_{c}$ vs $T/T_m$ obtained
from the velocity-force curves for the system in
Fig.~\ref{fig:5}.
(b) The corresponding $P_{6}$ vs $T/T_m$, showing
that the increase in $F_{c}$ is associated with a disordering transition.
NOTE THAT CHARLES SENT EXTRA IMAGES TO USE AS INSETS
}
\label{fig:6}
\end{figure}

From the velocity-force curves, we obtain 
the depinning force $F_{c}$,
defined as the value of $F_D$ at which we first observe
$\langle V_{||}\rangle>0.001$. 
In Fig.~\ref{fig:6} we plot $F_{c}$ and $P_{6}$ at depinning versus $T/T_m$, 
where $T_{m}$ is the temperature at which
the $F_{p} = 0$ system melts. Here, 
$F_{c}$ initially drops with increasing $T/T_m$ while
$P_{6}$ remains close to one, indicating that elastic depinning
is occurring. 
For temperatures above $T/T_m = 0.4346$,
there is a proliferation of topological defects
and $F_{c}$ first increases with temperature before decreasing
due to enhanced thermal hopping.

The overall behavior of $F_c$ shown in Fig.~\ref{fig:6} is
very similar to that found for the thermal peak
effect in type-II superconductors,
where there is a jump up in the critical
depinning force at the
elastic to plastic or order to disorder transition, coinciding with
a change in the shape of the
velocity-force curves
that causes a crossing of the curves
\cite{Kwok94,Higgins96,Paltiel00a,Ling01,Troyanovski02,Mohan07,ToftPetersen18,Tang96}
Figure~\ref{fig:6} also illustrates that the   
disordering transition occurs at $T/T_{m} < 1.0$,
similar to the peak
effect in superconductors
where the combination of thermal fluctuations and pinning
cause the vortex lattice to disorder
at temperatures below the clean melting temperature.
In general, for the superconducting system,
as $F_{p}$ increases, the temperature $T_p$ at which the peak effect occurs 
is reduced.   

\section{Dynamic Ordering}

In superconducting vortex systems, Wigner crystals,
and skyrmions, it was shown that when the
system is in a plastic depinning regime,
as the drive increases the effect of the
pinning on the moving particles is reduced
and the system
can dynamically order into a moving crystal
or moving smectic
with few topological defects at high drives
\cite{Bhattacharya93,Reichhardt17,Reichhardt22a,Reichhardt19b,Yaron95,Balents98,Pardo98,Olson98a,Reichhardt01}.
In Ref.~\cite{Bhattacharya93},
Higgins and Bhattacharya construct a dynamic
phase diagram across the peak
effect
showing a transition in the flow above depinning from elastic
to disordered,
where in the disordered regime above the peak effect, for higher
drives the 
flow becomes ordered again.
Koshelev and Vinokur \cite{Koshelev94}
then performed simulations of the plastic depinning of vortices, and
found that at higher drives when the vortices are moving
sufficiently rapidly over the pinning,
the vortex lattice can dynamically reform.
The drive at which this
dynamical ordering occurs diverges near
the melting temperature of
the pin-free system.

\begin{figure}
\includegraphics[width=3.5in]{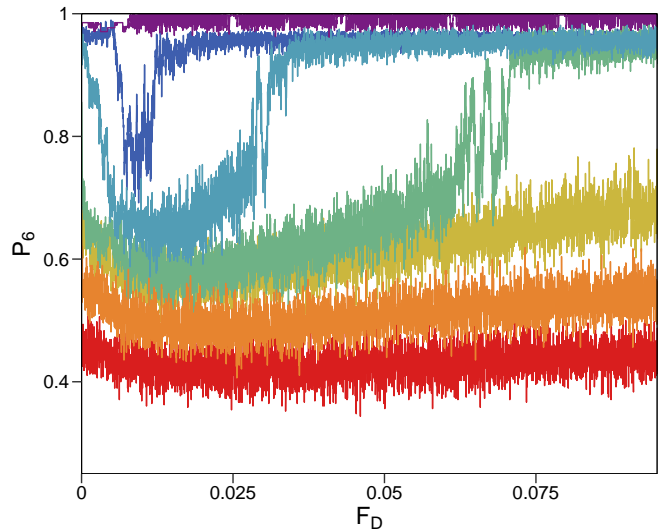}
\caption{
$P_{6}$ vs $F_{D}$ for the system in
Fig.~\ref{fig:5} for $T/T_m = 0.174$, 0.348, 0.608, 
0.869, 1.087, 1.304, and
$1.74$, from top to bottom.
For $T/T_m<1.0$, the system
dynamically orders into a moving crystal
at higher drives.
}
\label{fig:7}
\end{figure}

We next study whether dynamical reordering also occurs 
across the skyrmion peak effect
with characteristics that are similar to what is found for the
dynamical ordering across the peak effect in the
superconducting case \cite{Bhattacharya93,Koshelev94}. 
In Fig.~\ref{fig:7} we plot
$P_{6}$ versus $F_{D}$ for the system in
Figs.~\ref{fig:5} and \ref{fig:6}
at $T/T_m = 0.174$, 0.348, 0.608  0.869, 1.087, 1.304, and 
$1.74$.
At $T/T_m = 0.174$,
the system is always in an ordered state and $P_6 \approx 1$ at all drives,
while for $T/T_m=0.348$, 0.608, and 0.869,
the system is disordered at depinning
but at high drive undergoes
a dynamical reordering transition in which
$P_{6}$ increases
to a value close to $P_6=0.95$.
The drive at which the 
dynamical ordering occurs increases
with increasing $T/T_m$, and for $T/T_{m} > 1.0$,
the system remains disordered for all values
of $F_{D}$. 

\begin{figure}
\includegraphics[width=3.5in]{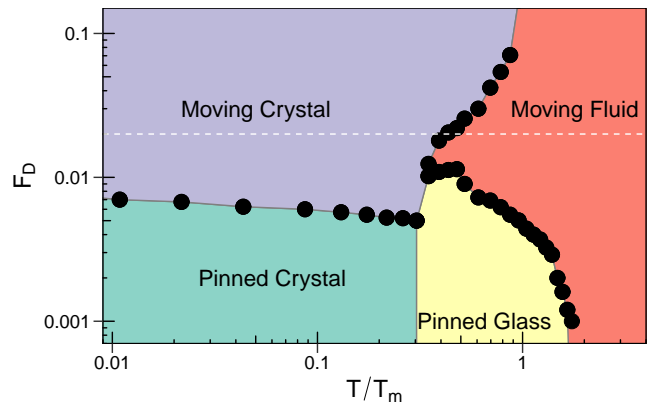}
\caption{
Dynamic phase diagram as a function of $F_{D}$ vs $T/T_m$
constructed from the velocity-force curves and $P_6$ data in  
Figs.~\ref{fig:6} and \ref{fig:8}
showing a pinned crystal, moving crystal,
pinned glass, and moving glass.
The horizontal dashed line indicates the value of $F_D$ at which
the velocity and skyrmion Hall angle data in Fig.~\ref{fig:9}
were obtained.
}
\label{fig:8}
\end{figure}

From the features of the $F_{c}$ and $P_{6}$
curves, we construct a dynamical phase
diagram as a function of $F_{D}$
versus $T/T_{m}$, as shown in Fig.~\ref{fig:8}.
There is a pinned
skyrmion crystal that depins elastically to a moving crystal as well as
a pinned glass that depins plastically to a moving fluid state.
Above the peak effect regime, there is a transition
from a moving fluid into a moving crystal with increasing drive,
and the value of $F_D$ at which this dynamic reordering occurs
diverges as $T/T_{m} = 1.0$ is approached.
For sufficiently high temperatures, the pinned
phase is lost and the system is permanently in the moving fluid state.
The overall shape of the phase diagram is very similar to
that of the
superconducting peak effect
dynamical phase diagram obtained experimentally by
Higgins and Bhattacharya \cite{Bhattacharya93},
while the divergence in the reordering driving force
upon approaching $T/T_m=1$ agrees with the simulation results
of Koshelev and Vinokur \cite{Koshelev94}.
We note that there could be
additional phases beyond those shown in Fig.~\ref{fig:8}.
For example, the moving crystal
could also contain regions of a moving smectic phase,
and it could be possible to draw a distinction
between a moving glass and a moving liquid.

\begin{figure}
\includegraphics[width=3.5in]{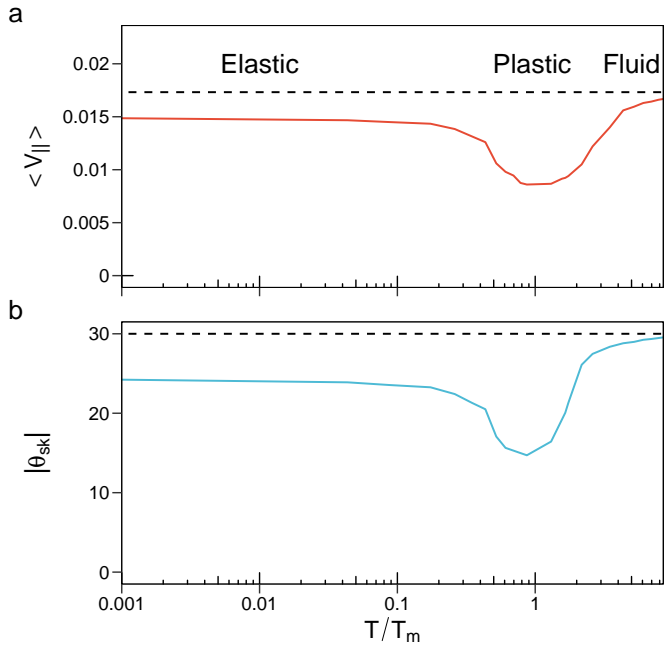}
\caption{(a) $\langle V_{||}\rangle$ vs $T/T_{m}$
for a slice at $F_D=0.02$, above the depinning force,
from the phase diagram in Fig.~\ref{fig:8}.
There is a dip across the
order to disorder transition.
(b) The corresponding skyrmion Hall angle
$|\theta_{sk}|$ vs $T/T_m$,
which also shows a dip.
For $T/T_{m} > 1.0$, the values of both quantities approach the
$F_p=0$ values (dashed lines).
}
\label{fig:9}
\end{figure}

In the superconducting system,
consequences of the peak effect appear even for drives well above
depinning. 
For example, it is possible to apply a constant current with
$F_{D} > F_{c}$  while increasing the temperature,
as indicated by the horizontal
dashed line at $F_D=0.2$ in Fig.~\ref{fig:8}.
Here,
the velocity parallel to the drive
exhibits
a drop at the temperature corresponding to
the elastic to plastic transition,
as shown in Fig.~\ref{fig:9}(a)
where there is
a minimum in $\langle V_{||}\rangle$ close to $T/T_m = 1.0$.
A similar dip appears in 
the perpendicular velocity and the absolute velocity.
As shown in Fig.~\ref{fig:9}(b),
there is a corresponding
drop in the magnitude of the skyrmion Hall angle.
For higher $T/T_m$, $|\theta_{sk}|$ approaches the clean limit
value expected in a sample with $F_p=0$.
The magnitude of the drop in $P_6$ diminishes
with increasing $F_{D}$.  

\begin{figure}
\includegraphics[width=3.5in]{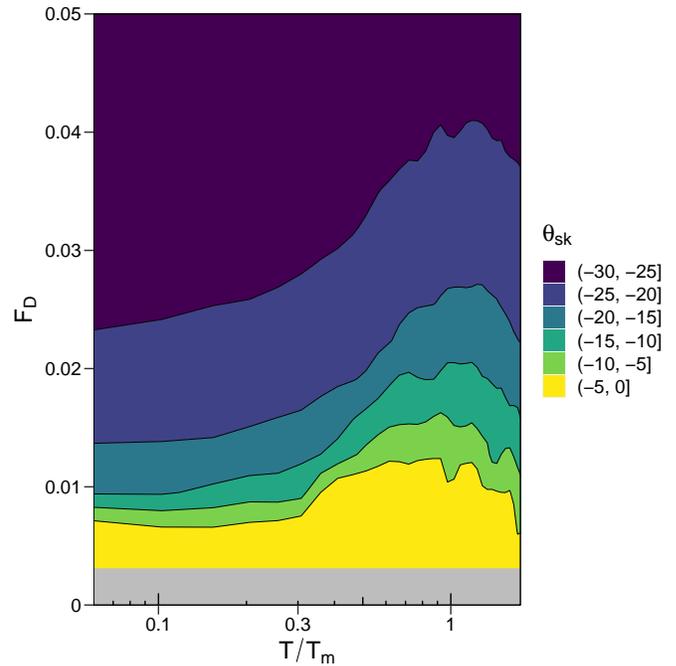}
\caption{Heat map of the skyrmion Hall angle $\theta_{sk}$ in degrees
as a function of $F_{D}$ vs $T/T_{m}$.  In the gray region,
the skyrmions are not flowing steadily and
$\theta_{sk}$ is undefined.  
}
\label{fig:10}
\end{figure}

In Fig.~\ref{fig:10} we show a heat map of the
skyrmion Hall angle
as a function of $F_D$ versus $T/T_m$.
For a fixed value of $F_D$,
$\theta_{sk}$ increases as the temperature passes through
the transition from elastic to plastic behavior.
The skyrmion Hall angle always increases with increasing $F_D$, and there
is also an increase in $\theta_{sk}$ with increasing temperature
for the highest values of $T/T_m$.

\begin{figure}
\includegraphics[width=3.5in]{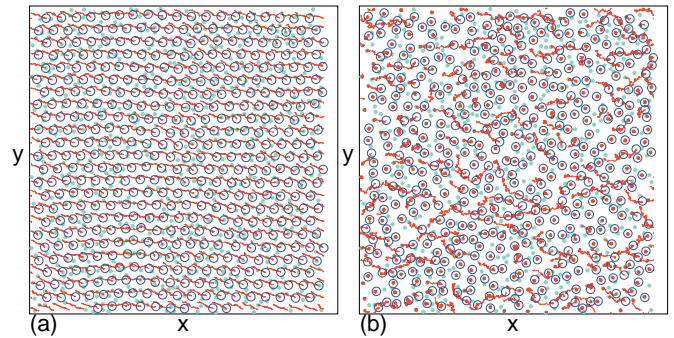}
\caption{The pinning site locations (filled circles), skyrmion
locations (open circles), and trajectories (lines)
for the system in Fig.~\ref{fig:8} at $F_{D}= 0.02$. 
(a) $T/T_m = 0.28$ where the flow is elastic.
(b) $T/T_m =1.74$ where the flow is fluid.
}
\label{fig:11}  
\end{figure}

In Fig.~\ref{fig:11}(a) we plot the
skyrmion locations and trajectories along with the pinning site
locations for the system in Fig.~\ref{fig:9}
at $F_{D}= 0.02$ and $T/T_m = 0.28$ in the 
elastic regime.
The skyrmions exhibit some random motion due to the thermal fluctuations,
but form a triangular lattice that moves as a rigid body at an angle to
the applied drive.
At $T/T_m=1.74$, Fig.~\ref{fig:11}(b)
indicates that the same system is disordered and contains regions of
flowing skyrmions coexisting with pinned regions.
Due to the significant amount of thermal hopping, finite skyrmion
velocities extend down to much lower values of $F_D$ compared
to the $T/T_m = 0.28$ sample.
At low drives in the elastic depinning regime,
the creep motion must occur in a correlated fashion since plastic
distortions of the skyrmion lattice are suppressed.
As a result, for $F_{D} = 0.02$ and $T/T_m = 0.28$, all of the skyrmions move
at the same velocity,
while for the
same drive at $T/T_m =1.74$,
due to the thermally induced breakdown of the lattice structure,
some skyrmions remain temporarily pinned while other skyrmions are
moving, giving a reduced average velocity.
This is what produces
the crossing of the velocity-force curves.
At even higher temperatures,
the hopping becomes so rapid that the
average velocity in the liquid phase at $F_{D} = 0.02$
is higher than the average velocity in the elastic regime, as was shown 
in Fig.~\ref{fig:9}.

\section{Skyrmion Density and Materials Parameters}

\begin{figure}
\includegraphics[width=3.5in]{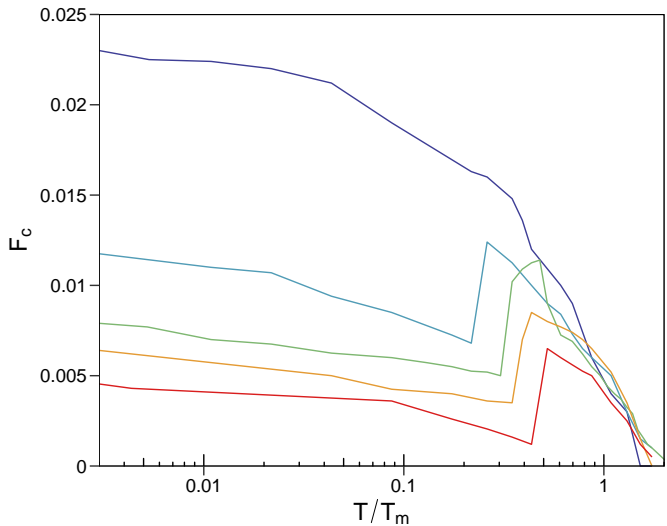}
\caption{ 
The critical depinning force $F_{c}$ vs $T/T_m$ for the system
in Fig.~\ref{fig:9} at varied skyrmion
density $n_s = 0.2$, 0.3, 0.4, 0.5, and $0.6$, from top to bottom.
For $n_s = 0.2$, the system is disordered even at
$T/T_m = 0$ and there is no peak effect.   
}
\label{fig:12}
\end{figure}

Since the peak effect is correlated with
an order to disorder transition, the next question 
is whether the peak effect can still occur if the system is always disordered.
In the case of a single skyrmion,
there should not be a peak effect since collective effects are
absent and there can be no
disorder to order transition.
This suggests that in samples with
finite pinning strength,
for low skyrmion densities when the skyrmions are sufficiently far apart that
skyrmion-skyrmion interactions become unimportant,
the skyrmion lattice
will always be disordered and the peak effect should be absent.  
In Fig.~\ref{fig:12} we plot the critical depinning force
$F_{c}$ versus $T/T_m$ for
skyrmion densities of $n_s = 0.2$, 0.3, 0.4, 0.5, and $0.6$.
Up until this point we have focused on samples with
$n_s = 0.4$.
For $n_s > 0.2$, the
skyrmions form a crystal and
undergo elastic depinning
over an extended  range of temperatures,
with a peak effect
appearing at the order to disorder transition
temperature.
For $n_s \leq 0.2$, the skyrmions
are sufficiently far apart that they
are disordered even at $T/T_m = 0$. In this case, the overall 
value of $F_{c}$ is much higher, and there is no peak effect.
For $n_s \geq 0.2$, as the skyrmion density 
increases, the peak effect shifts to higher temperatures
and the overall effectiveness of the pinning is reduced. 
This is consistent with observations in
superconductors where for cleaner samples the
overall pinning of the vortices is reduced and the peak effect
occurs at high temperatures, while for samples with strong pinning,
the peak effect is lost. 

\begin{figure}
\includegraphics[width=3.5in]{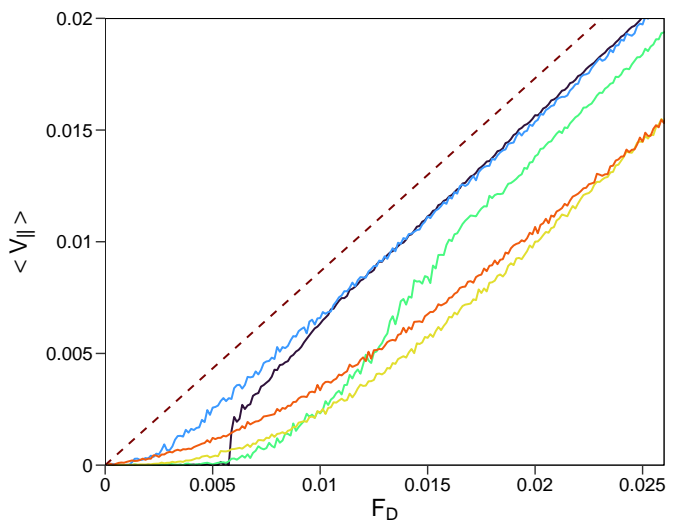}
\caption{ 
$\langle V_{||}\rangle$ vs $F_{D}$
for a sample from Fig.~\ref{fig:11} with  
$n_{s} = 0.6$
at $T = 0.0$ 
(black),
$0.5$ (blue),
$0.6$ (yellow),
$1.5$ (green),
and $2.0$ (orange).
There is a clear change in the shape of the velocity-force
curves across the peak effect, resulting in
a crossing of the velocity-force curves. 
}
\label{fig:13}
\end{figure}

As the skyrmion density increases,
there is a more pronounced difference in the
shape of the velocity-force curves
on either side of the peak effect,
as shown in Fig.~\ref{fig:13}
where we plot
$\langle V_{||}\rangle$ versus $F_{D}$
for a sample with $n_{s} = 0.6$ at
$T = 0$,
$0.5$,
$0.6$,
$1.5$,
and $2.0$.
The disordering
transition in this case occurs at $T = 0.6$, and the depinning
threshold reaches its highest value at this temperature.
For higher temperatures, the depinning threshold drops
and the velocity-force curves change shape,
resulting in a clear crossing of the curves.

\begin{figure}
\includegraphics[width=3.5in]{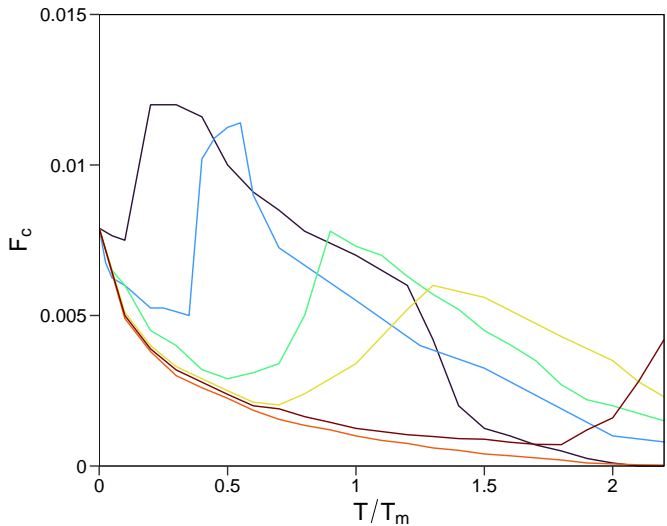}
\caption{$F_{c}$ vs $T/T_m$ for the same system
from Fig.~\ref{fig:9} with $n_s=0.4$
at $\kappa = 3$, 2, 1, 0.75, 0.5, and $0$ from
top to bottom. The peak effect shifts to higher
temperatures as $\kappa$ decreases. Here the curves have
been normalized to $T_m$ for the $\kappa=2.0$ system.
}
\label{fig:14}
\end{figure}

\begin{figure}
\includegraphics[width=3.5in]{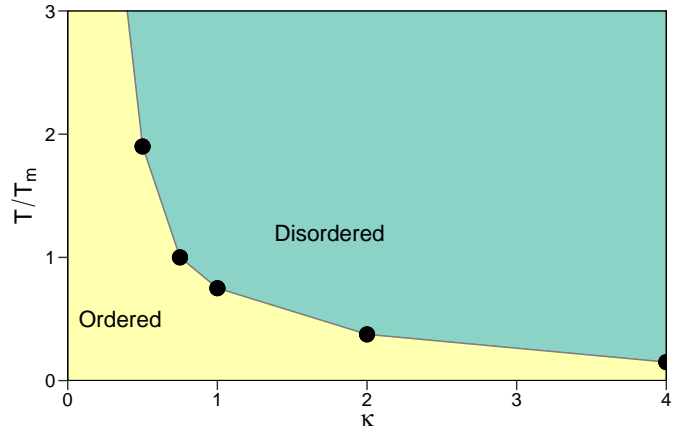}
\caption{Dynamic phase diagram of the order to
disorder transition as a function of $T$ vs $\kappa$
for the system in Fig.~\ref{fig:13} with $n_s=0.4$.
Here the temperature has been normalized using $T_m$
for the $\kappa=2.0$ system.
}
\label{fig:15}
\end{figure}

Since the temperature dependence of the skyrmion-skyrmion 
pairwise interaction is given by
$\exp(-\kappa T)$
in the model of
Wang {\it et al.} \cite{Wang22a}, we next consider 
how the peak effect
varies for changing $\kappa$.
In Fig.~\ref{fig:14} we plot $F_{c}$ versus $T/T_m$ for
the $n_s=0.6$ sample from Fig.~\ref{fig:9} at
$\kappa = 3$, 2, 1, 0.75, 0.5, and $0$.
For $\kappa = 3.0$,
the peak effect appears at low temperatures.
The $\kappa = 2.0$ system was discussed above,
and
for $\kappa = 1.0$, 0.75, and $0.5$, the
peak effect occurs at higher temperatures
while the decrease in $F_{c}$ with increasing $T/T_m$
becomes more pronounced as $\kappa$ gets smaller.
For $\kappa = 0.0$, there
is no peak effect and
$F_{c}$ drops all the way to zero before the system can disorder.
In this case, by the time the
temperature is high enough to melt the
skyrmion lattice, the individual skyrmions
are easily able to thermally hop
out of the pinning wells.
This result indicates that
for large $\kappa$ the system may disorder rapidly with
increasing $T$ or always remain disordered, while 
in samples with small $\kappa$,
the peak effect will occur at higher temperatures
closer to the bulk melting 
transition.
In Fig.~\ref{fig:15} we highlight
a dynamic phase diagram as a function of
$T/T_m$ versus $\kappa$ showing the order to disorder transition.
The transition point should also depend
on the strength of the pinning and the skyrmion density.

\section{Discussion}

An open question is whether a skyrmion peak effect
could also occur as function of applied magnetic field. 
If the strength of the skyrmion-skyrmion interaction diminishes
with increasing field, it could be possible that an
order to disorder transition accompanied
by a peak effect could occur for increasing magnetic field. 
In superconducting vortex systems,
the number of vortices increases
monotonically with magnetic field, so the elastic modulus of the vortex
lattice rises with increasing magnetic field,
but near the upper critical
field $H_{c2}$,
the vortex-vortex interaction
strength is reduced when the penetration depth diverges. 
For skyrmions, the magnetic field dependence
is more complex, since not only is the number of skyrmions not a
monotonic function of applied magnetic field, but also the size of the
skyrmions can change with magnetic field.
In principle,
a peak effect will occur whenever the
skyrmions form an elastic lattice that can  transition into 
a disordered state as a function of
applied magnetic field or temperature.
There is already some evidence that a skyrmion 
lattice can thermally melt \cite{Balaz21},
so it would be interesting to study the transport
in these systems
while some feature such as skyrmion density or pinning strength is tuned.

In this work, we considered 2D systems; however,
many skyrmion systems contain three-dimensional (3D) line-like objects,
and a peak effect is known to occur in superconducting vortex systems
where the vortices take the form of 3D lines.
As long as the strength of the pairwise skyrmion interactions
decreases with increasing temperature,
there should be a peak effect in a 3D skyrmion system if pinning
is present.
In fact, it could possible
that the order-disorder transition at the peak effect
is second order in 2D systems and first order in 3D systems.
If a skyrmion peak effect can occur in 3D samples,
it could have a variety of interesting properties
such as metastable dynamics.
These could appear
when the system is prepared in a disordered state under conditions where
the steady state is ordered, or vice versa.
Metastable behaviors have been observed in superconducting vortices
near the peak effect regime \cite{Henderson96,Andrei06}.
We used a particle-based model to explore the skyrmion peak
effect; however, it would be interesting to further probe the behavior
with a continuum-based model, provided that the system is sufficiently
large to exhibit a clear elastic to plastic transition.
The continuum model could could capture
additional effects such as shape distortions
of the skyrmions across the peak effect. 
We also assumed a constant Magnus force; however, if the skyrmions
change in size across the peak effect, the Magnus force could 
be altered.

\section{Summary}
We have numerically investigated a thermal skyrmion peak effect
for a skyrmion lattice driven over quenched disorder.
As the thermal fluctuations increase,
the effectiveness of the pinning is reduced due to thermal
hopping, but
the strength of the skyrmion-skyrmion
interactions is exponentially reduced
as proposed by Wang {\it et al.} \cite{Wang22a}. 
The skyrmion peak effect is similar to the peak effect 
associated with an order to disorder or elastic to plastic transition  
in superconducting vortex systems.
When the pinning is weak, the
skyrmions form a lattice that depins elastically at $T = 0$.
As the temperature increases,
the depinning threshold decreases,
but when the skyrmion-skyrmion interactions become soft enough,
there is an order to disorder transition to a skyrmion
glass phase accompanied by a large increase in
the depinning threshold.
As the temperature is further increased,
thermal hopping reduces the depinning threshold again, resulting in
the appearance of
a peak in the depinning force. The peak effect
occurs for temperatures below the melting temperature of the skyrmion
lattice in the absence of quenched disorder.
Across the peak effect, we find
a change in the shape of the
velocity-force curves and a crossing of the curves similar to what
is found for
the superconducting peak effect. 
For a constant drive,
even above the depinning threshold
the velocity and skyrmion Hall angle both show a drop across the
peak effect.
We construct a dynamical phase diagram
and find that there can be a dynamical reordering transition
for increasing drive above the peak effect temperature
but below the clean melting temperature.
We map the different phases and the nonmonotonic behavior
of the skyrmion Hall angle through the peak effect.
We show that the
peak effect should be robust
whenever a skyrmion lattice can form with skyrmion-skyrmion interactions that
are modified by temperature.
Our results are also consistent with several experimental studies that
have shown an increase
in the critical current for skyrmion lattices with increasing temperature. 

\acknowledgments
We gratefully acknowledge the support of the U.S. Department of Energy
through the LANL/LDRD program for this work.
This work was supported by the US Department of Energy through the
Los Alamos National Laboratory. Los Alamos National Laboratory is operated
by Triad National Security, LLC, for the National Nuclear Security
Administration of the U. S. Department of Energy (Contract No.
892333218NCA000001).

\bibliography{mybib}

\end{document}